%% file: HIG-17-019_temp.tex
\pdfoutput=1

\documentclass[11pt,twoside,a4paper,cmspaper,final,collab]{cms-tdr}
\def\svnVersion{486256P}\def\cmsCernNoTag{CERN-EP-2018-165}\def\cmsCernDate{\today}\def\cmsMessage{Published in Physical Review Letters as \href{http://dx.doi.org/10.1103/PhysRevLett.122.021801}{\doi{10.1103/PhysRevLett.122.021801.}}}

\begin{document}\cmsNoteHeader{HIG-17-019}

\hyphenation{had-ron-i-za-tion}
\hyphenation{cal-or-i-me-ter}
\hyphenation{de-vices}
\RCS$HeadURL: svn+ssh://svn.cern.ch/reps/tdr2/papers/HIG-17-019/trunk/HIG-17-019.tex $
\RCS$Id: HIG-17-019.tex 479502 2018-10-28 12:20:39Z alverson $

\newcolumntype{x}[1]{D{.}{.}{#1}}
\newcommand{\mc}[1]{\multicolumn{1}{c}{#1}}

\newlength\cmsFigWidth
\newlength\cmsFigWidthLarge

\newlength\cmsTabSkip\setlength{\cmsTabSkip}{1ex}

\ifthenelse{\boolean{cms@external}}{\setlength\cmsFigWidth{0.98\columnwidth}}{\setlength\cmsFigWidth{0.48\textwidth}}
\ifthenelse{\boolean{cms@external}}{\setlength\cmsFigWidthLarge{0.98\columnwidth}}{\setlength\cmsFigWidthLarge{0.6\textwidth}}
\ifthenelse{\boolean{cms@external}}{\providecommand{\cmsTabSpace}{\hspace{-2ex}}}{\providecommand{\cmsTabSpace}{}}
\providecommand{\cmsTable}[1]{\resizebox{\textwidth}{!}{#1}}

\newcommand{\mmm}{\ensuremath{m_{\Pgm\Pgm}}}
\newcommand{\mH}{\ensuremath{m_\PH}}
\newcommand{\Htomm}{\ensuremath{\PH\!\to\!\Pgmp\Pgmm}}

\cmsNoteHeader{HIG-17-019}
\title{Search for the Higgs boson decaying to two muons in proton-proton collisions at \texorpdfstring{$\sqrt{s}=13\TeV$}{sqrt(s) = 13 TeV}}

\date{\today}

\abstract{ A search for the Higgs boson decaying to two oppositely charged
muons is presented using data recorded by the CMS experiment at the CERN LHC in
2016 at a center-of-mass energy $\sqrt{s}=13\TeV$, corresponding to
an integrated luminosity of 35.9\fbinv. Data are found to be compatible with
the predicted background. For a Higgs boson with a mass of
125.09\GeV, the 95\% confidence level observed (background-only expected) upper limit on the
production cross section times branching fraction to a pair of muons is found
to be 3.0 (2.5) times the standard model expectation.
In combination with data recorded at center-of-mass energies $\sqrt{s}=7$ and
8\TeV, the background-only expected upper limit improves to 2.2 times the
standard model value with a standard model expected significance of 1.0 standard deviations. The
corresponding observed upper limit is 2.9 with an observed significance of 0.9
standard deviations.  This corresponds to an observed upper limit on the
standard model Higgs boson branching fraction to muons of $6.4\times10^{-4}$
and to an observed signal strength of $1.0\pm{1.0}\stat\pm0.1\syst$.  }

\hypersetup{
	pdfauthor={CMS Collaboration},
pdftitle={Search for the Higgs boson decaying to two muons in proton-proton collisions at sqrt(s)=13 TeV},
pdfsubject={CMS},
pdfkeywords={CMS, Higgs, Muon, Couplings, Second-generation, Rare-decays}}

\maketitle

In the standard model (SM), the masses of fermions are generated by their
Yukawa coupling to the Higgs
field~\cite{Englert:1964et,Higgs:1964ia,Higgs:1964pj,Higgs:1966ev}, whose
existence was confirmed by the Higgs boson (\PH)
discovery~\cite{Aad:2012tfa,Chatrchyan:2012xdj,Chatrchyan:2013lba}.
Measurements at CMS and ATLAS provided evidence that the Higgs boson couples to
bottom quarks~\cite{CMS-HIG-16-044,Aaboud2017}, and established that it couples
with tau leptons~\cite{CMS-HIG-16-043,Aad2015} and top
quarks~\cite{HIG-17-035,CERN-EP-2018-138}. The Higgs boson mass has been
measured and is found to be
$\mH=125.09\pm0.24\GeV$ in a combination of ATLAS and CMS
data samples~\cite{Aad:2015zhl}. The study of the Higgs boson decays
to muons is of particular importance because it extends the investigation to
its couplings to fermions of the second generation.  For a Higgs boson with a
mass of 125.09\GeV, the expected branching fraction ($\mathcal{B}$) to muons is
$2.17 \times 10^{-4}$~\cite{LHCHXSWG4}, and the narrow decay width of the Higgs
boson \cite{Dittmaier:2011ti,Khachatryan:2014iha} is several orders of
magnitude smaller than the $\mathcal{O}(\GeV)$ experimental dimuon mass
resolution.  The signal would appear as a narrow resonance over a smoothly
falling mass spectrum from the SM background processes, primarily Drell--Yan
(DY) and leptonic \ttbar decays.

The CMS and ATLAS Collaborations placed upper limits on the product of the
Higgs boson production cross section and branching fraction
$\mathcal{B}(\Htomm)$ of approximately seven times the SM value at 95\%
confidence level (\CL) with LHC Run~1
data~\cite{Aad:2014xva,Khachatryan:2014aep}, collected at center-of-mass
energies $\sqrt{s}=7$ and 8\TeV. The ATLAS Collaboration improved its expected
limit to 2.9 times the SM expectation by adding 36.1\fbinv{} of data collected
at 13\TeV~\cite{Aaboud:2017ojs} and measured an observed limit of 2.8 times the
SM expectation. This Letter presents a search for $\Htomm$ events with the CMS
detector  using 35.9\fbinv of proton-proton ($\Pp\Pp$) collision data collected
in 2016 at $\sqrt{s}=13\TeV$, and its combination with the data collected at
$\sqrt{s} = 7$ and 8\TeV corresponding to integrated luminosities of 5.0\fbinv
and 19.7\fbinv, respectively.

The central feature of the CMS apparatus is a superconducting solenoid of
$6\unit{m}$ internal diameter, providing a magnetic field of $3.8\unit{T}$.
Within the solenoid volume are a silicon pixel and strip tracker, a lead
tungstate crystal electromagnetic calorimeter, and a brass and scintillator
hadron calorimeter, each composed of a barrel and two endcap sections. Forward
calorimeters extend the pseudorapidity ($\eta$) coverage provided by the barrel
and endcap detectors. Muons are detected in gas-ionization detectors embedded
in the steel flux-return yoke outside the solenoid. A more detailed description
of the CMS detector, together with a definition of the coordinate system used
and the relevant kinematic variables, can be found in
Ref.~\cite{Chatrchyan:2008zzk}.

The Monte Carlo (MC) simulated events used to model the signal include the four
leading Higgs boson production processes: gluon-gluon fusion ($\Pg\Pg\PH$),
vector boson fusion (VBF), and associated production with a vector boson
(V$\PH$, V=$\PW$, or $\cPZ$) or top quarks ($\ttbar\PH$). The Higgs boson MC
samples are generated at next-to-leading order (NLO) for masses of 120, 125,
and 130\GeV with \POWHEG 2.0~\cite{Nason:2004rx}, using the parton distribution
function sets of NNPDF3.0~\cite{Ball:2014uwa}. The $\Pg\Pg\PH$ acceptance in
each analysis category is found to be in agreement with that
calculated at NLO with the \MGvATNLO{2.2.2}~\cite{Alwall:2014hca} generator.
The SM background processes considered are DY, single and pair production of
top quarks (s\cPqt, \ttbar), and di- and triboson production (VV, VVV).
Simulated background processes are used only to optimize the event selection
and not for the final background estimate, which is obtained from data.
Background samples are generated using \MGvATNLO and \POWHEG. Spin
correlations in multiboson processes generated using \MGvATNLO are simulated using
\textsc{MadSpin}~\cite{Artoisenet:2012st}. The parton shower and hadronization
processes are modeled by the  {{\PYTHIA}8.212}~\cite{Sjostrand:2007gs}
generator with the CUETP8M1~\cite{Khachatryan:2015pea} underlying event tune.
The detector response is based on a detailed description of the CMS detector
and is simulated with the \GEANTfour package~\cite{Agostinelli:2002hh}.
Simultaneous $\Pp\Pp$ interactions overlapping the event of interest (pileup)
are included in the simulated samples. The distribution of the number of
additional interactions per bunch crossing in the simulation corresponds to that
observed in the 13\TeV data collected in 2016, with an average of 23
interactions. The SM Higgs boson cross section and branching fractions are
taken from the LHC Higgs boson cross section working group recommendations
\cite{LHCHXSWG4}, while cross sections for the background processes are taken
from  \FEWZ{3.1}~\cite{FEWZ}, \textsc{Top++2.0}~\cite{TOPPP},
\textsc{hathor}~\cite{HATHOR2,HATHOR1}, and \MCFM~\cite{Campbell2011}.

The particle-flow (PF) algorithm~\cite{Sirunyan:2017ulk} is used to reconstruct
observable particles in each event. It combines all subdetector information to
reconstruct individual particles and identify them as charged or neutral
hadrons, photons, or leptons. Electron and muon candidates are formed by associating a
track in the silicon detectors with a cluster of energy in the electromagnetic
calorimeter~\cite{Khachatryan:2015hwa} or a track in the muon system. The
relative transverse momentum (\pt) resolution of muon candidates with
$\pt<100\GeV$ is $\lesssim$1.6\% in the barrel \cite{MUO-16-001}.  Jets are
reconstructed using the anti-\kt clustering algorithm~\cite{Cacciari:2008gp}
with a distance parameter of 0.4, as implemented in the \FASTJET
package~\cite{Cacciari:2011ma}. Jets are required to have a minimum \pt of
30\GeV and a maximum $\abs{\eta}$ of 4.7. Further identification criteria are
applied in order to reject jets from pileup  or noise present in the detector
\cite{CMS-PAS-JME-16-003}. For jets with $\abs{\eta} < 2.4$, multivariate
algorithms discriminate jets arising from the hadronization of
b-quarks~\cite{BTV-16-002}. The missing transverse momentum $\ptmiss$ is
defined as the magnitude of the negative vector \ptvec sum of all reconstructed
particles (charged and neutral) in the event and is modified by corrections to
the energy scale of reconstructed jets.  The reconstructed vertex with the
largest value of summed physics-object $\PT^2$ is taken to be the primary $\Pp\Pp$
interaction vertex.

Events are selected by the trigger system requiring the presence of at least
one isolated muon with $\pt>24\GeV$~\cite{cmstrigger}. The offline selection,
optimized to maximize the sensitivity of the analysis,
requires at least two oppositely charged muons with $\pt > 26 \GeV$  ($\pt > 20
\GeV$) for the leading (subleading) muon and $\abs{\eta}< 2.4$. To
reject events with muons from nonprompt decays, muons must be isolated, with a
relative isolation sum $<$25\%. The relative isolation sum is
calculated as the scalar \pt sum of PF objects, excluding the muon, within a
cone of radius $\Delta R = \sqrt{\smash[b]{(\Delta \eta)^2+(\Delta \phi)^2}}=
0.4$ centered on the direction of the muon, and divided by the muon \pt.
Charged particles not associated with the event vertex are not considered in
this sum, and a correction is applied in order to account for the neutral
particle contamination arising from pileup~\cite{IsolationPU}. The invariant
mass of the Higgs boson candidate ($\mmm$) is constructed from the two highest
\pt oppositely charged muons, and the event is retained for further analysis if
$110 < \mmm < 150\GeV$. The overall trigger efficiency for these events is 98.5\%.

\begin{figure}[htb]
	\centering
	\includegraphics[width=\cmsFigWidth]{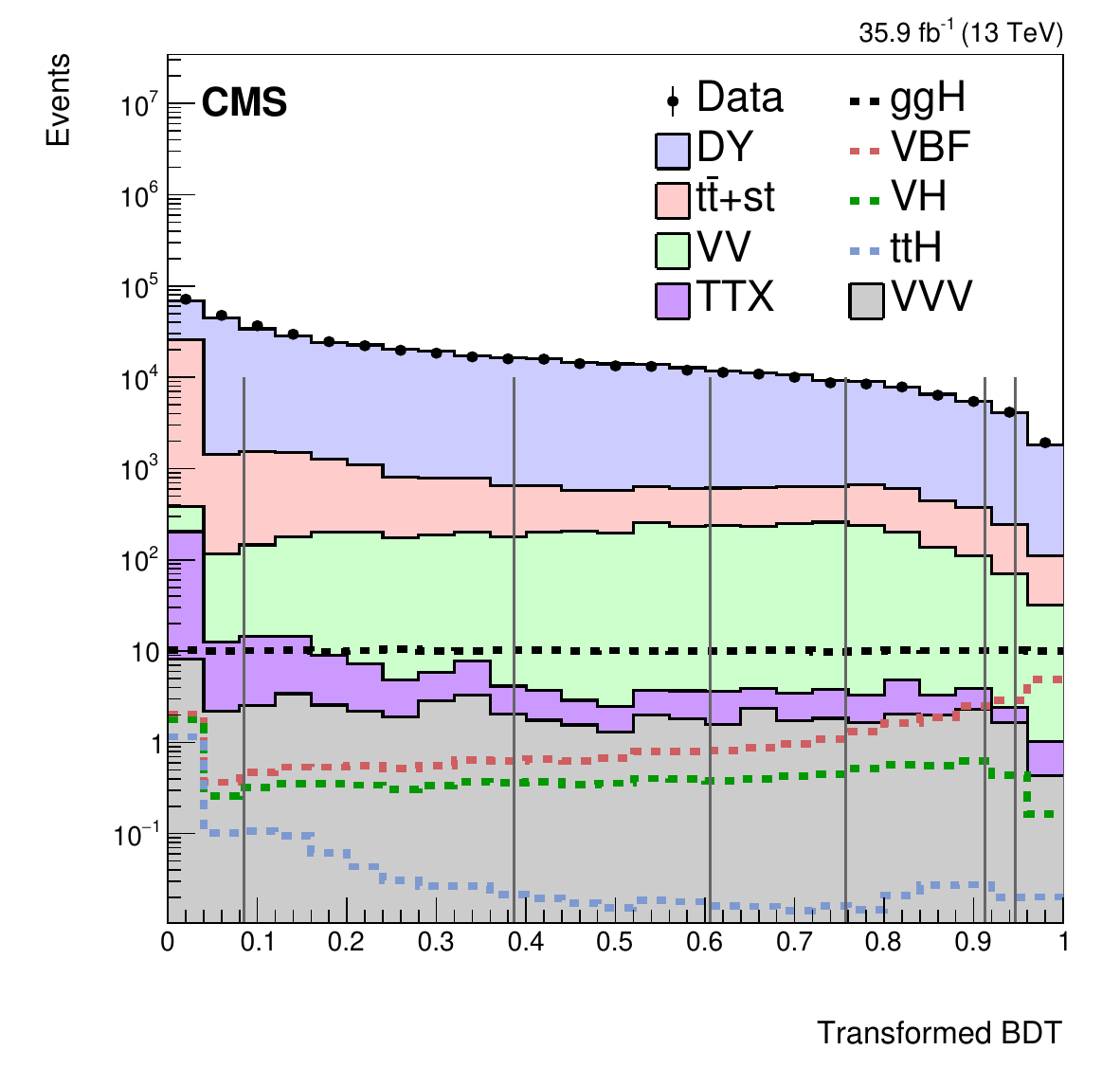}
	\caption{The transformed BDT output distributions in data (solid points) and MC simulation (histograms).
	The stacked solid histograms represent the background processes, while the stacked dashed histograms represent the signal.
	In the legend, V denotes the vector bosons $\PW$ and $\PZ$,
    and TTX indicates the top quark pair production in association with a vector boson V or another top quark pair.
    The vertical lines denote the BDT response intervals indicated in Table~1.}
	\label{fig:spectrum}
\end{figure}

Events are classified into categories using variables that are largely
uncorrelated with $\mmm$ in order to enhance the sensitivity to the Higgs boson
signal. The primary Higgs boson production mechanisms targeted by this analysis
are VBF and $\Pg\Pg\PH$. The \pt and $\eta$ of the dimuon system, and the
$\abs{\Delta\eta}$ and $\abs{\Delta\phi}$ between the muons, distinguish between
$\Pg\Pg\PH$ signal events and the DY background.  The $\abs{\eta}$ of each of the
two highest \pt jets, the mass and $\abs{\Delta\eta}$ between the jets in each of
the two highest mass dijet pairs, and the number of jets with $\abs{\eta} <
2.4$ (central jets) and $\abs{\eta} > 2.4$ (forward jets) identify VBF signal
events.  Finally, the number of b-tagged jets and $\ptmiss$ identify events
with \ttbar decays.  These variables are used as input to a boosted decision
tree (BDT)~\cite{Hocker:2007ht}, which was trained with simulated signal and background
events normalized to their respective SM cross sections.
The dimuon mass and its resolution are not used as input to the BDT in order to avoid biasing the background shape,
but are used in the signal extraction as discussed later.
Simulated signal events used in the training
steps are not used later in the analysis.  Figure~\ref{fig:spectrum} shows the
BDT output distributions for data and for simulated events. The output of the
classifier was transformed such that the sum of all signal events has a uniform
distribution. A large fraction of the VBF signal events can be distinguished
from background processes, and corresponds to events with the highest BDT
score.

The event categories are defined using the BDT score and the expected dimuon
mass resolution, gauged by the largest $\abs{\eta}$ of the two muons.
The best mass resolution is obtained when both muons are located in the central part of the
detector $\abs{\eta}< 0.9$, where the muon momentum resolution is approximately
constant, and degrades when one of the muons is more forward, especially in the
region $\abs{\eta}> 1.9$, where there are reduced lever arm
and increased multiple scattering within the tracking volume.

The number of categories and the values of the BDT and $\abs{\eta}$ boundaries of the
categories were optimized according to an iterative process using
$\sum_{i} S^{2}_{i}/B_{i}$ as a figure of merit, where $S_{i}$ and $B_{i}$ are
the number of expected signal and background events in each category in the
$i^{\text{th}}$ mass bin from 120 to 130\GeV with 0.5\GeV spacing.  A first
category boundary is created by optimizing the figure of merit against all
possible boundaries in $\abs{\eta}$ and in BDT score separately, and then choosing the
one with the larger gain. The process is then repeated recursively within each
of the two newly created categories to create additional category boundaries
within them until a set number of categories is achieved. Some rounding of the
values of the boundaries was made afterward, checking that the simplification
does not significantly worsen the expected limit.

This procedure incorporates the dimuon mass
resolution into the definition of the categories, optimizing the sensitivity of the
analysis.  This optimization results in 15 categories shown in
Table~\ref{tab:nEvts}. Simulated events are used to optimize the event
categories and to estimate the selection efficiency for signal events. In each
category, the shape and the normalization of the dimuon mass distribution of
the background contributions are obtained from a parametric fit to the data
using a set of empirical functions. The product of signal acceptance and
efficiency for the $\Htomm$ signal varies depending on production process.  This
product is shown in Table~\ref{tab:nEvts} for each category for a Higgs boson
mass of 125\GeV, together with the functional form used to derive the
background from data and the $S/\sqrt{B}$ ratio within the full
width at half maximum (FWHM) of the expected signal distribution.

The reconstructed invariant mass of the signal is modeled with a sum of up to
three Gaussian functions, which provides a satisfactory description of the
low-mass tail of the distribution and each model is separately fit to the simulated
dimuon invariant mass distribution for each production process in each category
for $\mH= 120$, 125, and 130\GeV. The fit parameters are interpolated for
masses within that range.  The invariant mass distribution of the background
primarily follows the smoothly falling spectrum of the high-mass DY background.
Secondary contributions come from the single and pair production of top quarks.
In each category, the background distribution is modeled by fitting the data
with a single analytic function, chosen from a set of alternative options.
These include a sum of exponential functions, Bernstein polynomials
$(B_{\text{deg\,n}})$, and a modified version of the Breit--Wigner $\cPZ$ boson
line shape derived and validated by fitting \FEWZ predictions of the DY
invariant mass distribution at
next-to-NLO~\cite{Bourilkov:2016qum,Bourilkov:2016oet}:

\begin{equation}
	\text{mBW}(x)=\frac{\re^{\,a_{2}x+a_{3}x^{2}}} {(x-{m}_{\cPZ})^{a_{1}}+(\frac{\Gamma_{\cPZ}}{2})^{a_{1}}},
\end{equation}
where ${m}_{\cPZ}$ and $\Gamma_{\cPZ}$ are the mass and the width of the {\cPZ} boson fixed to known values~\cite{pdg}.
In addition, \FEWZ spectra templates multiplied by polynomial functions
are considered, as well as a modified Breit--Wigner distribution
multiplied by a Bernstein polynomial of up to degree 4 ($B_{\text{deg\,4}}$).
The chosen function maximizes the expected sensitivity
while introducing only a negligible
bias in the measured signal yield, which is determined
as follows.
In each category, background-only fits to the data are performed with every function.
From each of these fits, thousands of pseudo-data sets are generated,
taking into account the uncertainties in the fit parameters and their
correlations, and simulated signal events are added according to their expected
SM yields.  Each of the background functions is then used to fit the
pseudo-data sets generated from every other function, with the
total signal yield floating freely in the fit.  The bias is
estimated as the median excess or deficit in the measured signal yield
relative to the SM expectation.
Accepted functions in each category have a maximum possible bias of less
than 20\% of the statistical uncertainties for $\mH= 120$, 125,
and 130\GeV.
Including these deviations as spurious signals leads to an overall uncertainty
in the calculated limit of less than 1\%, which is neglected.
Correlation between bias terms is also found to be negligible.

{\setlength{\extrarowheight}{1.2pt}
\begin{table*}[!hbtp]
	\centering
	\topcaption{ \label{tab:nEvts}
	The optimized event categories, the product of acceptance and selection efficiency in \% for the different production processes, the total expected number of SM signal events ($\mH=125\GeV$), the estimated number of background events per GeV at 125\GeV, the FWHM of the signal peak, the background functional fit form, and the $S/\sqrt{\smash[b]{B}}$ ratio within the FWHM of the expected signal distribution.
	}
	\setlength\tabcolsep{4pt}

    \cmsTable{
    \begin{scotch}{ r@{ -- }l r@{$\abs{\eta}$}l *{8}{x{-1}} c x{-1} }
		\multicolumn{2}{c}{BDT response} & \multicolumn{2}{c}{Maximum}       & \mc{$\Pg\Pg\PH$}  & \mc{VBF}  & \mc{$\PW\PH$} & \mc{$\cPZ\PH$} & \mc{$\ttbar\PH$}
		& \mc{Signal}  & \mc{$\mathrm{Bkg}/{\GeVns}$} & \mc{FWHM}           & \mc{Bkg fit} & \mc{$S/\sqrt{\smash[b]{B}}$} \\ [1.2pt]
        \multicolumn{2}{c}{quantile [\%]}     & \multicolumn{2}{c}{muon $\abs{\eta}$} & \mc{[\%]} & \mc{[\%]} & \mc{[\%]}   & \mc{[\%]}    & \mc{[\%]}
		&              & \mc{$@ 125\GeV$}          & \mc{$[{\GeVns}]$} & \mc{function} & \mc{@ FWHM}       \\ [2pt]
		\hline
		$ 0 $ & $  8$ &         & $ < 2.4$  &  4.9 &  1.3 &  3.3 &  6.3 &   \mc{32} &  21.2 &  \mc{$3.13\times10^{3}$} & 4.2 & \mc{mBW $B_{\text{deg\,4}}$} & 0.12\\
		$ 8 $ & $ 39$ & $1.9 <$ & $ < 2.4$  &  5.6 &  1.7 &  3.9 &  3.5 &  1.3 &  22.3 &  \mc{$1.34\times10^{3}$} & 7.2 & \mc{mBW $B_{\text{deg\,4}}$} & 0.16\\
		$ 8 $ & $ 39$ & $0.9 <$ & $ < 1.9$  & \mc{10} &  2.8 &  6.5 &  6.4 &  5.2 &  41.1 &  \mc{$2.24\times10^{3}$} & 4.1 & \mc{mBW $B_{\text{deg\,4}}$} & 0.29\\
		$ 8 $ & $ 39$ &         & $ < 0.9$  &  3.2 &  0.8 &  1.9 &  2.1 &  3.5 &  12.7 &  \mc{$ 7.83\times10^{2}$} & 2.9 & \mc{mBW $B_{\text{deg\,4}}$} & 0.18\\
		$39 $ & $ 61$ & $1.9 <$ & $ < 2.4$  &  2.9 &  1.7 &  2.7 &  2.7 &  0.3 &  11.8 &  \mc{$ 4.37\times10^{2}$} & 7.0 & \mc{mBW $B_{\text{deg\,4}}$} & 0.14\\
		$39 $ & $ 61$ & $0.9 <$ & $ < 1.9$  &  7.2 &  3.3 &  6.1 &  5.2 &  1.3 &  29.2 &  \mc{$ 9.70\times10^{2}$} & 4.0 & \mc{mBW $B_{\text{deg\,4}}$} & 0.31\\
		$39 $ & $ 61$ & 	& $ < 0.9$  &  3.6 &  1.1 &  2.6 &  2.2 &  0.9 &  14.5 &  \mc{$ 4.81\times10^{2}$} & 2.8 & \mc{mBW} & 0.26\\
		$61 $ & $ 76$ & $1.9<$  & $ < 2.4$  &  1.2 &  1.5 &  1.8 &  1.7 &  0.2 &   5.2 &  \mc{$ 1.48\times10^{2}$} & 7.6 & \mc{mBW $B_{\text{deg\,4}}$} & 0.11\\
		$61 $ & $ 76$ & $0.9 <$ & $ < 1.9$  &  4.8 &  3.6 &  4.5 &  4.4 &  0.7 &  20.3 &  \mc{$ 5.12\times10^{2}$} & 4.2 & \mc{mBW $B_{\text{deg\,4}}$} & 0.29\\
		$61 $ & $ 76$ &         & $ < 0.9$  &  3.2 &  1.6 &  2.3 &  2.1 &  0.6 &  13.1 &  \mc{$ 3.22\times10^{2}$} & 3.0 & \mc{mBW} & 0.28\\
		$76 $ & $ 91$ & $1.9 <$ & $ < 2.4$  &  1.2 &  3.1 &  2.2 &  2.1 &  0.2 &   5.8 &  \mc{$ 1.04\times10^{2}$} & 7.1 & \mc{mBW $B_{\text{deg\,4}}$} & 0.14\\
		$76 $ & $ 91$ & $0.9 <$ & $ < 1.9$  &  4.4 &  8.7 &  6.2 &  6.0 &  1.1 &  20.3 &  \mc{$ 3.60\times10^{2}$} & 4.2 & \mc{mBW $B_{\text{deg\,4}}$} & 0.35\\
		$76 $ & $ 91$ &         & $ < 0.9$  &  3.1 &  4.0 &  3.8 &  3.6 &  0.9 &  13.7 &  \mc{$ 2.36\times10^{2}$} & 3.2 & \mc{mBW} & 0.34\\
		$91 $ & $ 95$ &         & $ < 2.4$  &  1.7 &  6.4 &  2.5 &  2.6 &  0.5 &   8.6 &  96.0 & 4.0 & \mc{mBW} & 0.28\\
		~~~~$95 $ & $100$ &         & $ < 2.4$  &  2.0 & \mc{19} &  1.5 &  1.4 &  0.7 &  13.7 &    83.4 & 4.1 & \mc{mBW} & 0.48\\[\cmsTabSkip]
        \multicolumn{2}{c}{Total} &         & $ < 2.4$  & \mc{59} & \mc{61} & \mc{51} & \mc{52} & \mc{49} & \mc{253} & \mc{$1.30\times10^{4}$} & 3.9 & & \\
	\end{scotch}
    } 
\end{table*}}

The systematic uncertainties considered in the analysis account for possible
mismodeling in the signal shape or rate.  The shape of the reconstructed Higgs
boson invariant mass is affected by the muon momentum scale and resolution.
Uncertainties in the calibration of these values are propagated to the shape of
the invariant mass distribution of the Higgs boson, assuming a Gaussian prior,
yielding variations of up to 0.05\% in the position of the peak and up to 10\%
in its width. Jet energy uncertainties in scale and resolution affect the
analysis through migrations between categories. The largest variation of this
kind amounts to 6\% of the relative yield.  Uncertainty in the simulation of
additional pileup events is modeled by varying the total inelastic cross
section~\cite{Aaboud:2016mmw,cms:2018inelasticxsec} by $\pm5\%$, which
translates to $\approx$1\% variations in the yields. The systematic
uncertainty in the b-tagging or light-quark and gluon jets mistagging
efficiencies result in event migration across categories of $\approx$1\%.
Lepton efficiency mismodeling is accounted for with trigger and isolated muon
identification uncertainties ($\approx$2\%). The factorization and
renormalization scales used in the MC simulations are varied up and down
separately by a factor of 2, translating to changes of up to 6\% in the signal
acceptance per category. The parton distribution functions used in the signal
MC simulations are varied using the NNPDF3.0 replicas, which yield differences
of $\approx$2\%. In the comparison of measured signal yields with
expectation, additional uncertainties in the calculated signal cross sections
are considered. They are due to the choice of factorization and renormalization
scale (3.9, 0.4, 3.8, 1.9, and $\lesssim$10\%, for $\Pg\Pg\PH$, VBF, $\PZ\PH$,
$\PW\PH$, and $\ttbar\PH$, respectively) and parton distribution functions
(3.2, 2.1, 1.6, 1.9, and 3.7\%), as well as the 1.7\% uncertainty in the
$\Htomm$ branching fraction \cite{LHCHXSWG4}. Finally, a 2.5\% uncertainty is
associated with the integrated luminosity measurement~\cite{LUM-17-001}.

\begin{figure}[htb] \centering
	\includegraphics[width=\cmsFigWidthLarge]{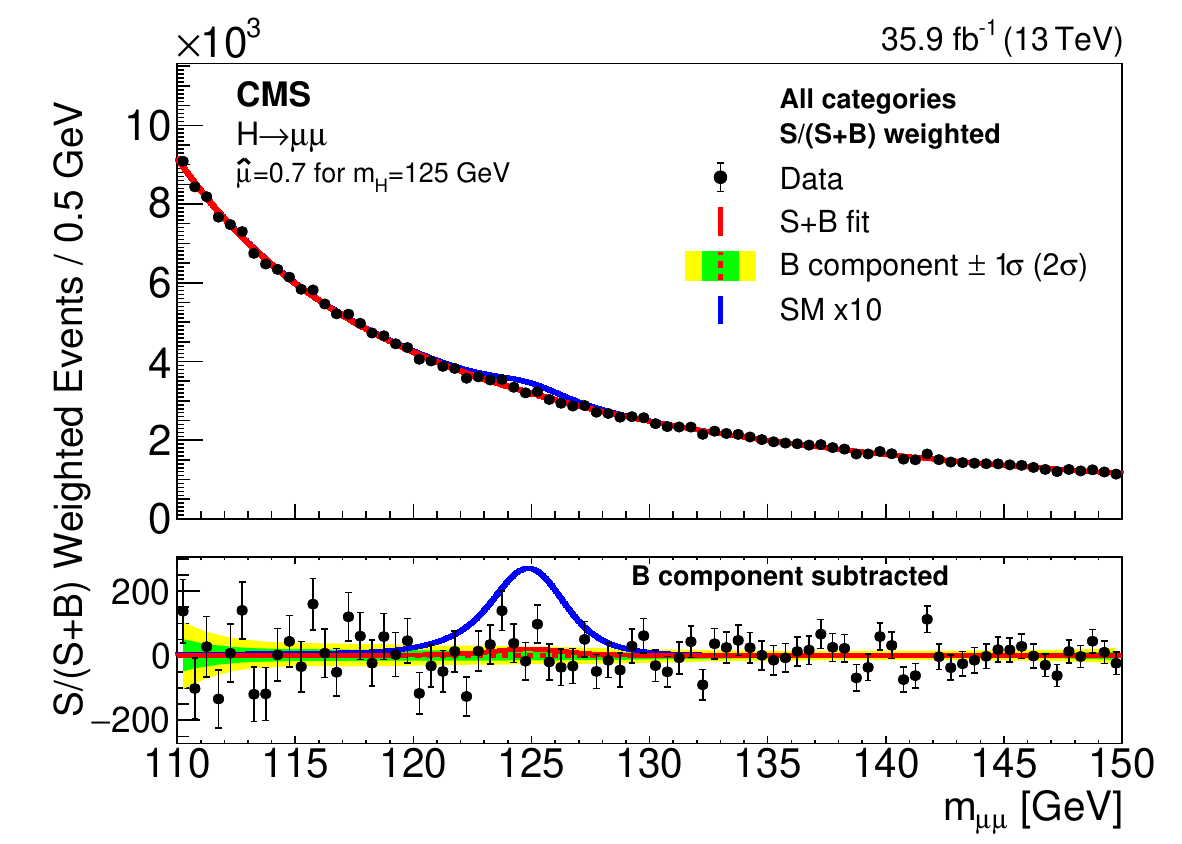}
    \caption{Data and weighted sum of signal-plus-background fits to each category.
	Events are weighted according to the expected ratio of signal to signal-plus-background
	in the category to which they belong.
    The solid blue line represents the expectation from the SM Higgs boson multiplied by a factor of 10.
    The lower panel shows the difference between the data and the background component of the fit.
    }
	\label{fig:invmass} \end{figure}

A maximum likelihood signal-plus-background fit to the dimuon invariant mass
spectrum is performed across all categories to measure the signal strength
modifier $\mu$, defined as $(\sigma\mathcal{B})_\textup{obs} /$
$(\sigma\mathcal{B})_\textup{SM}$ where $\sigma$ indicates the Higgs boson
production cross section.  The best fit signal strength for a Higgs boson mass
hypothesis of 125.09\GeV ($\hat{\mu}_{125}$) and 68\% \CL interval are extracted
with a profile likelihood ratio, according to the procedure described in
Ref.~\cite{Khachatryan:2014jba}, yielding $\hat{\mu}_{125}=0.7\pm
1.0\stat^{+0.2}_{-0.1}\syst$ for $\mH=125.09\GeV$~\cite{COMB-NOTE}.
Figure~\ref{fig:invmass} shows the background component and the
signal-plus-background fits to the data in all categories combined, weighted by
the expected ratio of signal to signal-plus-background in each category. The 95\% \CL upper
limit on the signal strength modifier computed with the asymptotic \CLs
method~\cite{Junk,Read,CLs} and the compatibility of the dimuon yield with the
background-only hypothesis for the 2016 data set (13\TeV) are also derived. The
observed (expected for $\mu=0$) upper limit at 95\% \CL for $\mH=125.09\GeV$ is
3.0 (2.5), with an observed (expected for $\mu=1$) significance of the
incompatibility with the background-only hypothesis of 0.6 (0.9) standard
deviations (s.d.).

\begin{figure}[htb] \centering
	\includegraphics[width=\cmsFigWidthLarge]{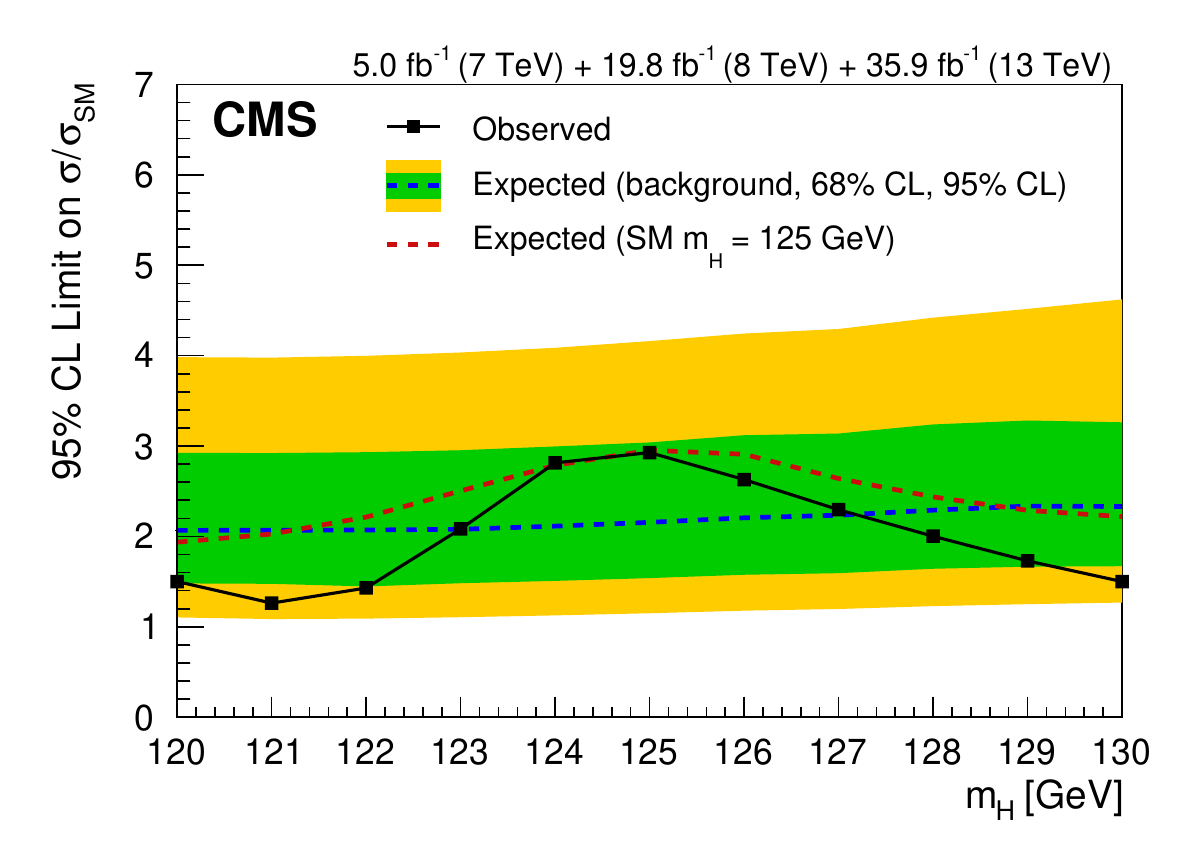} \caption{The
	95\% \CL upper limit on the signal strength modifier, $\mu$,
        in the region around the Higgs boson mass for the
	combination of the 7, 8, and 13\TeV data sets together with the
	expected limit obtained in the background only hypothesis (dashed
	black line) and in the signal-plus-background hypothesis (dashed red
	line) for the SM Higgs boson with $\mH=125\GeV$.  } \label{fig:limits}
\end{figure}

The 95\% \CL upper limit on the signal strength as a function of $\mH$
in the region around the Higgs boson mass for a
combination of data recorded at center-of-mass energies of 7, 8, and 13\TeV is
shown in Fig.~\ref{fig:limits}, and yields an observed (expected for $\mu=0$)
limit on the production rate of 2.9 (2.2) times the SM value at
$\mH=125.09\GeV$. The observed limit generally agrees well with
the expected limit curve for $\mu=1$ that is also shown, and
corresponds to an upper limit on the $\Htomm$ branching
fraction of $6.4\times10^{-4}$, assuming the SM production cross sections. The
best fit signal strength for $\mH=125.09\GeV$ is
$\hat{\mu}_{125}^\textup{comb}=1.0 \pm 1.0\stat^{+0.1}_{-0.1}\syst$, and the
observed combined significance is 0.9 s.d.  The expected value
for $\mu=1.0$ is $\hat{\mu}_{125}^\textup{comb}=1.0^{+1.1}_{-1.0}$ and the
combined expected significance is 1.0 s.d. Theoretical
uncertainties are considered correlated across the data sets, while the main
experimental uncertainties are considered uncorrelated.

In summary, we present a search for the Higgs boson decaying to two muons using
data recorded by the CMS experiment at the LHC in 2016 at a center-of-mass
energy of 13\TeV, corresponding to an integrated luminosity of 35.9\fbinv.  No
significant evidence for this decay is observed.  Limits are set on the cross
section times branching fraction of the Higgs boson decaying to two muons. The
combination with data recorded at center-of-mass energies of 7 and 8\TeV yields
a 95\% confidence level observed upper limit of 2.9 times the standard model
value for $\mH=125.09\GeV$.
The corresponding expected upper limit in the absence
of a SM decay in this channel is 2.2, which is the most sensitive to date.
This corresponds to an observed (standard model expected) significance of the Higgs boson
decaying into two muons of 0.9 (1.0) standard deviations and
an observed signal strength of $1.0\pm1.0\stat\pm0.1\syst$.
Assuming standard model production cross sections for the Higgs boson, the observed limit corresponds
to an upper limit of $6.4\times10^{-4}$ on the Higgs boson branching fraction
to two muons.

\begin{acknowledgments}
	We congratulate our colleagues in the CERN accelerator departments for the excellent performance of the LHC and thank the technical and administrative staffs at CERN and at other CMS institutes for their contributions to the success of the CMS effort. In addition, we gratefully acknowledge the computing centers and personnel of the Worldwide LHC Computing Grid for delivering so effectively the computing infrastructure essential to our analyses. Finally, we acknowledge the enduring support for the construction and operation of the LHC and the CMS detector provided by the following funding agencies: BMWFW and FWF (Austria); FNRS and FWO (Belgium); CNPq, CAPES, FAPERJ, and FAPESP (Brazil); MES (Bulgaria); CERN; CAS, MoST, and NSFC (China); COLCIENCIAS (Colombia); MSES and CSF (Croatia); RPF (Cyprus); SENESCYT (Ecuador); MoER, ERC IUT, and ERDF (Estonia); Academy of Finland, MEC, and HIP (Finland); CEA and CNRS/IN2P3 (France); BMBF, DFG, and HGF (Germany); GSRT (Greece); NKFIA (Hungary); DAE and DST (India); IPM (Iran); SFI (Ireland); INFN (Italy); MSIP and NRF (Republic of Korea); LAS (Lithuania); MOE and UM (Malaysia); BUAP, CINVESTAV, CONACYT, LNS, SEP, and UASLP-FAI (Mexico); MBIE (New Zealand); PAEC (Pakistan); MSHE and NSC (Poland); FCT (Portugal); JINR (Dubna); MON, RosAtom, RAS and RFBR (Russia); MESTD (Serbia); SEIDI, CPAN, PCTI and FEDER (Spain); Swiss Funding Agencies (Switzerland); MST (Taipei); ThEPCenter, IPST, STAR, and NSTDA (Thailand); TUBITAK and TAEK (Turkey); NASU and SFFR (Ukraine); STFC (United Kingdom); DOE and NSF (USA).
\end{acknowledgments}

\bibliography{auto_generated}
\cleardoublepage \appendix\section{The CMS Collaboration \label{app:collab}}\begin{sloppypar}\hyphenpenalty=5000\widowpenalty=500\clubpenalty=5000\input{HIG-17-019-authorlist.tex}\end{sloppypar}
\end{document}

%% file: HIG-17-019-authorlist.tex
\vskip\cmsinstskip
\textbf{Yerevan Physics Institute, Yerevan, Armenia}\\*[0pt]
A.M.~Sirunyan, A.~Tumasyan
\vskip\cmsinstskip
\textbf{Institut f\"{u}r Hochenergiephysik, Wien, Austria}\\*[0pt]
W.~Adam, F.~Ambrogi, E.~Asilar, T.~Bergauer, J.~Brandstetter, E.~Brondolin, M.~Dragicevic, J.~Er\"{o}, A.~Escalante~Del~Valle, M.~Flechl, R.~Fr\"{u}hwirth\cmsAuthorMark{1}, V.M.~Ghete, J.~Hrubec, M.~Jeitler\cmsAuthorMark{1}, N.~Krammer, I.~Kr\"{a}tschmer, D.~Liko, T.~Madlener, I.~Mikulec, N.~Rad, H.~Rohringer, J.~Schieck\cmsAuthorMark{1}, R.~Sch\"{o}fbeck, M.~Spanring, D.~Spitzbart, A.~Taurok, W.~Waltenberger, J.~Wittmann, C.-E.~Wulz\cmsAuthorMark{1}, M.~Zarucki
% ...
\vskip\cmsinstskip
% ...
1:  Also at Vienna University of Technology, Vienna, Austria\\
% ...
